\documentclass[twocolumn,showpacs,preprintnumbers,amsmath,amssymb]{revtex4}
\usepackage{amssymb}
\usepackage[dvips]{graphicx}
\begin{document}
\title{Checkerboard density of states in strong-coupling
superconductors}
\author{A.S.~Alexandrov}

\affiliation{
Department of Physics, Loughborough University,Loughborough LE11\,3TU, UK\\
}

\begin{abstract}

The Bogoliubov-de Gennes (BdG) equations are solved in the
strong-coupling limit, where real-space (preformed) pairs
bose-condense with finite center-of-mass momenta. There are two
energy scales in this regime, a temperature independent incoherent
gap $\Delta_p$ and a temperature dependent coherent gap $\Delta_c
(T)$, modulated in real space. The single-particle density of
states (DOS) reveals checkerboard   modulations similar to the
tunnelling DOS in cuprates.
\end{abstract}

\pacs{PACS:  74.72.-h, 74.20.Mn, 74.20.Rp, 74.25.Dw}
\vskip2pc]

\maketitle

Many independent observations  show that the superconducting state
of cuprates is as anomalous as the normal state. In particular,
there is strong evidence for a $d$-like order parameter, which
changes sign when the $CuO_{2}$ plane is rotated by $\pi /2$
\cite{ann}. A few  phase-sensitive experiments \cite{pha} provide
unambiguous evidence in this direction.    A $d$-wave BCS gap
could appear in the two-dimensional Hubbard model near half
filling, as suggested by Scalapino, Loh, and Hirsch \cite{sca2}
concurrently with the discovery of novel superconductors. On the
other hand,   $c$-axis Josephson tunnelling \cite{klem},
high-precision magnetic measurements \cite {mulsym}, photo-excited
quasi-particle relaxation dynamics \cite{kab} and some other
experiments \cite{guomeng} support
 more conventional anisotropic $s$-like  gap.

 In fact, there are stronger
  deviations from the conventional Fermi/BCS-liquid behaviour than the gap symmetry.
 There is now convincing evidence  for  pairing of carriers  well above T$_c$ as predicted
 by the bipolaron theory
 \cite{ale},  the  clearest one is provided by the uniform magnetic susceptibility
\cite{alekab,mul},  tunnelling, and photoemission. The tunnelling
and photoemission gap is
 almost temperature independent below T$_{c}$ \cite{brus1,ren} and exists  above
T$_{c}$ \cite{ren,din,she} with its maximum  several times larger
than  expected in the weak and  intermediate-coupling \cite{eli}
BCS theory. Kinetic \cite{bat} and thermodynamic \cite{lor} data
suggest that the gap
 opens both in  charge and spin channels  at any relevant
temperature in a wide range of doping. A plausible explanation is
that the normal state (pseudo)gap, $\Delta_p$, is half of the
bipolaron binding energy \cite{alegap}, although alternative
models have been proposed \cite{tim}.

Further studies of the gap function revealed even more
 complicated physics.
 Reflection experiments, in which an incoming electron from the
normal
 side of a normal/superconducting contact is reflected as a hole along
 the same trajectory \cite{and}, showed a much smaller  gap edge than the
 bias  at the tunnelling conductance maxima
 \cite{yag}.  Two distinctly
 different gaps with different magnetic field and temperature
 dependence were observed in the c-axis I(V) characteristics  \cite{kras}.
   They were also
observed   with the femtosecond time-resolved optical spectroscopy
\cite{kab2}. More recent STM experiments revealed
 checkerboard spatial modulations of the tunnelling DOS,
  with \cite{hoff} and without \cite{hoff2,kapit} applied
magnetic fields.

We have proposed a simple  model \cite{aleand} explaining two
different gaps in  cuprates. The main assumption, supported by a
parameter-free estimate  of the Fermi energy \cite{alefermi}, is
 that the attractive potential is large compared with the renormalised Fermi
 energy, so that the ground state is the Bose-Einstein condensate of
 tightly bound real-space pairs. In this letter I calculate the single particle DOS
 of  strong-coupling (bosonic) superconductors
 by solving the inhomogeneous BdG equations.
When pairs are Bose-condensed  with finite center-of-mass momenta,
I obtain a checkerboard DOS reminiscent of  the tunnelling DOS in
cuprates.

The anomalous Bogoliubov-Gor'kov average
\[
F_{ss^{\prime }}({\bf r}_{1},{\bf r}_{2})= \left\langle \hat
{\Psi} _{s}({\bf r}_{1})\hat{\Psi}_
{s^{\prime }}({\bf
r}_{2})\right\rangle  ,
\]
is the superconducting order parameter both in the weak and
strong-coupling
regimes. It depends on the relative coordinate ${\bf \rho =r}_{1}-{\bf r}%
_{2} $ of two electrons (holes), described by field operators $\hat{\Psi} _{s}({\bf r})$,
 and on the center-of-mass coordinate $%
{\bf R}=({\bf r}_{1}+{\bf r}_{2})/2$. Its Fourier transform, $f({\bf %
k,K})$, depends on the relative momentum ${\bf k}$ and on the
center-of-mass momentum ${\bf K.}$ In the BCS theory ${\bf K=}0$,
and the Fourier transform of the order parameter is
proportional to the gap in the quasi-particle excitation spectrum, $f({\bf k,K%
})\varpropto \Delta _{{\bf k}}$. Hence the symmetry of the order
parameter and the symmetry of the gap are the same in the
weak-coupling regime. Under the rotation of the coordinate system, $\Delta _{%
{\bf k}}$ changes its sign, if the Cooper pairing appears in the
d-channel. The Cooper pairing might also take place with finite
center-of-mass momentum, if electrons are spin polarized
\cite{lar}.

On the other hand,  the symmetry of the order parameter could be
different from the `internal' symmetry of the pair wave-function,
and from the symmetry of a single-particle excitation gap in the
strong-coupling regime \cite{ale}. Real-space pairs might have an
unconventional symmetry due to a specific symmetry of the pairing
potential as in the case of the Cooper pairs \cite{sca2}. The
d-wave symmetry of the ground state could be also due to a
topological
 degeneracy of inter-site pairs on a square lattice
\cite{alekor}, as proposed in Ref. \cite{and2}. But in any case
the ground state and DOS are homogeneous, if pairs are condensed
with ${\bf K}=0$. However, if a pair band dispersion has its
minima at finite ${\bf K}$ in the center-of-mass Brillouin zone,
the Bose condensate is inhomogeneous. In particular,   the
center-of-mass bipolaron energy bands could have their minima at
the Brillouin zone boundaries at ${\bf K}=(\pi,0)$ and  three
other equivalent momenta \cite{alesym} (here and further I take
the lattice constant $a=1$, and $\hbar=1$). These four states are
degenerate, so that
the condensate wave function $\psi({\bf m})$ in the real (Wannier) space, ${\bf m}%
=(m_{x},m_{y}),$ is their superposition,
\begin{equation}
\psi({\bf m})=\sum_{{\bf K}=(\pm \pi ,0),(0,\pm \pi )}b_{{\bf K}%
}e^{-i{\bf K\cdot m}},
\end{equation}
where $b_{{\bf K}}=\pm \sqrt{n_{c}}/2$ are $c$-numbers,  and
$n_c(T)$ is the  atomic density of the Bose-condensate. The
superposition, Eq.(1), respects the time-reversal and parity
symmetries, if
\begin{equation}
\psi ({\bf m})=\sqrt{n_{c}}\left[ \cos (\pi m_{x})\pm \cos (\pi
m_{y})\right] .
\end{equation}
Two order parameters, Eq.(2), are physically identical because
they are related by the translation transformation. They have
$d$-wave symmetry  changing sign in the real space, when the
lattice is rotated by $\pi /2$. This symmetry is
entirely due to the pair-band energy dispersion with four minima at ${\bf %
K} \neq 0$, rather than due a  specific pairing potential. It
reveals itself as a {\it checkerboard} modulation of the hole
density with two-dimensional patterns, oriented along the
diagonals.  From this insight  one can expect  a fundamental
connection between stripes detected by different techniques
\cite{tran,bia}  and the symmetry of the order parameter in
cuprates \cite{alesym}.

Now  let us take into account that in the superconducting state
($T<T_c$)  single-particle excitations
 interact with the pair condensate via the same short-range attractive potential, which forms the pairs \cite{aleand}.
 The Hamiltonian  describing
the interaction of  excitations with the pair Bose-condensate in
the Wannier representation is
\begin{eqnarray}
 H &=& -\sum_{s,{\bf m,n}}[t({\bf m-n})+\mu \delta_{\bf m,n}]c^{\dagger}_{s \bf m}c_{s \bf
 n}\cr
 &+& \sum_{\bf m}[\Delta({\bf m})c^{\dagger}_{\uparrow \bf m}c_{\downarrow \bf
 m}+H.c.],
 \end{eqnarray}
 where $s= \uparrow,\downarrow$ is the
 spin, $t({\bf m})$ and $\mu$ are  hopping integrals and the chemical potential, respectively,
 $c^{\dagger}_{s \bf m}$ and $c_{s \bf
 m}$ create (annihilate) an electron  or hole at site ${\bf m}$, and $\Delta({\bf m}) \propto \psi ({\bf
 m})$. Applying  equations of motion for  the Heisenberg operators
 $\tilde{c}^{\dagger}_{s \bf m}(t)$ and $\tilde{c}_{s \bf
 m}(t)$, and the Bogoliubov transformation \cite{bog}
 \begin{equation}
\tilde{c}_{\uparrow \bf m}(t)=\sum_{\nu} [u_{\nu}({\bf m})
\alpha_{\nu} e^{-i\epsilon_{\nu}t} +v_{\nu}^* ({\bf
m})\beta_{\nu}^{\dagger} e^{i\epsilon_{\nu}t}],
\end{equation}
\begin{equation}
\tilde{c}_{\downarrow \bf m}(t)=\sum_{\nu} [u_{\nu}({\bf m})
\beta_{\nu} e^{-i\epsilon_{\nu}t} - v_{\nu}^* ({\bf
m})\alpha_{\nu}^{\dagger} e^{i\epsilon_{\nu}t}],
\end{equation}
one  obtains BdG equations describing  the single-particle
excitation spectrum,
\begin{equation}
\epsilon u({\bf m})=-\sum_{\bf n} [t({\bf m-n})+\mu \delta_{\bf
m,n}]u({\bf n}) + \Delta({\bf m})v ({\bf m}),
\end{equation}
\begin{equation}
-\epsilon v({\bf m})=-\sum_{\bf n} [t({\bf m-n})+\mu \delta_{\bf
m,n}]v({\bf n}) + \Delta({\bf m})u({\bf m}),
\end{equation}
where  excitation quantum numbers $\nu$ are omitted for
transparency. These equations are supplemented by the sum rule
$\sum_{\nu} [u_{\nu}({\bf m})u_{\nu}^*({\bf n}) + v_{\nu}({\bf
m})v_{\nu}^*({\bf n})]=\delta_{\bf m,n}$, which provides the Fermi
statistics of single particle excitations $\alpha$ and $\beta$.
Different from the conventional BdG equations in the weak-coupling
limit, there is virtually no feedback  of  single particle
excitations on the off-diagonal potential, $\Delta({\bf m})$, in
the strong-coupling regime. The number of these excitations is low
at temperatures below $\Delta_p/k_B$, so that the coherent
potential $\Delta ({\bf m})$ is an external (rather than a
self-consistent) field,  solely determined by the pair Bose
condensate \cite{aleand}.

While the analytical solution is not possible for any arbitrary
off-diagonal interaction $\Delta({\bf m})$, one can readily solve
the infinite system of discrete equations (6,7) for a periodic
$\Delta({\bf m})$ with a period commensurate with the lattice
constant, for example
\begin{equation}
\Delta({\bf m})= \Delta_c [e^{i\pi m_x} - e^{i\pi m_y}],
\end{equation}
which corresponds to the pair condensate at ${\bf K}=(\pm\pi,0)$
and $(0,\pm\pi)$, Eq.(2), with a temperature dependent (coherent)
$\Delta_c \propto \sqrt {n_c(T)}$. In this case the quasi-momentum
${\bf k}$ is the proper quantum number, $\nu= {\bf k}$, and the
excitation wave-function is a superposition of plane waves,
\begin{eqnarray}
u_{\nu}({\bf m}) &=&u_{{\bf k}}e^{i{\bf k}\cdot {\bf m}}+\tilde{u}_{{\bf k}}e^{i({\bf k-g})\cdot {\bf m}}, \\
v_{{\nu}}({\bf m}) &=&v_{{\bf k}}e^{i{({\bf k-g}_x})\cdot {\bf
m}}+\tilde{v}_{{\bf k}}e^{i({{\bf k-g}_y})\cdot {\bf m}}.
\end{eqnarray}
Here ${\bf g}_x=(\pi,0)$, ${\bf g}_y=(0,\pi)$, and ${\bf
g}=(\pi,\pi)$ are reciprocal doubled lattice vectors. Substituting
Eqs.(9) and (10) into Eqs.(6,7) one obtains four coupled algebraic
equations,
\begin{eqnarray}
\epsilon _{\bf k}u_{\bf k}&=&\xi_{\bf k}u_{\bf k}-\Delta_c (v_{\bf k}-\tilde{v}_{\bf k}), \\
\epsilon _{\bf k}\tilde{u}_{\bf k}&=&\xi_{\bf k-g}\tilde{u}_{\bf
k}+\Delta_c (v_{\bf k}-\tilde{v}_{\bf k}),
\\
-\epsilon_{\bf k}v_{\bf k}&=&\xi_{{\bf k-g}_x}v_{\bf k}+\Delta_c
(u_{\bf k}-\tilde{u}_{\bf k}), \\ -\epsilon_{\bf k}\tilde{v}_{\bf
k}&=&\xi_{{\bf k-g}_y}\tilde{v}_{\bf k}-\Delta_c (u_{\bf
k}-\tilde{u}_{\bf k}) ,
\end{eqnarray}
where $ \xi_{\bf k}=-\sum_{\bf n}t({\bf n}) e^{i{\bf k \cdot n}}
-\mu$. The determinant of the system (11-14) yields the following
equation for  the energy spectrum $\epsilon$:
\begin{eqnarray}
&&(\epsilon-\xi_{\bf k})(\epsilon-\xi_{\bf
k-g})(\epsilon+\xi_{{\bf k-g}_x})(\epsilon+\xi_{{\bf k-g}_y})\cr
&=&\Delta_c^2(2\epsilon+\xi_{{\bf k-g}_x}+\xi_{{\bf
k-g}_y})(2\epsilon-\xi_{\bf k}-\xi_{\bf k-g}).
\end{eqnarray}
Two positive  roots for $\epsilon$ describe the single-particle
excitation spectrum. Their calculation is rather cumbersome, but
not in the extreme strong-coupling limit, where the pair binding
energy $2\Delta_p$ is large compared with $\Delta_c$ and with the
single-particle bandwidth $w$. The
 chemical potential in this limit is pinned
 below a single-particle band edge, so that $\mu=-(\Delta_p+w/2)$ is negative, and
its magnitude is large compared with $\Delta_c$.
 Then the right
hand side in Eq.(15) is a perturbation, and the spectrum is
\begin{eqnarray}
\epsilon_{1\bf k}&\approx& \xi_{\bf k} -{\Delta_c^2\over{\mu}}, \\
\epsilon_{2\bf k}&\approx& \xi_{\bf k-g} -{\Delta_c^2\over{\mu}},
\end{eqnarray}
Its dispersion along the diagonal direction is shown in Fig.1 in
the nearest neighbor approximation for the hopping integrals on a
square lattice.

\begin{figure}
\begin{center}
\includegraphics[angle=-00,width=0.40\textwidth]{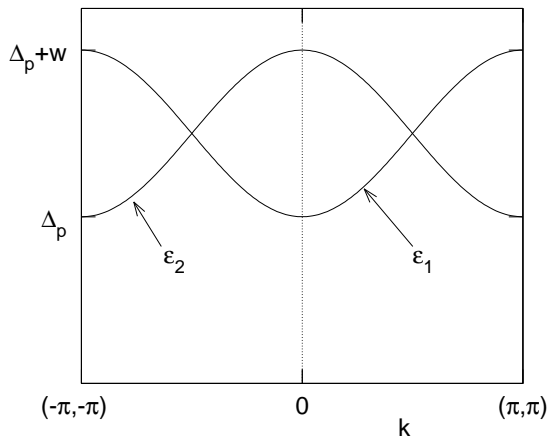}
 \caption{Single-particle excitation energy spectrum ( arb. units)
 along the diagonal direction of the two-dimensional Brillouin zone
  }
\end{center}
\end{figure}
If a metallic tip is placed at the point ${\bf m}$ above  the
surface of a sample, the STM current $I(V, {\bf m)}$ creates an
electron (or hole) at this point. Applying the Fermi-Dirac golden
rule and the Bogoliubov transformation, Eq.(4,5), and assuming
that the temperature is much lower than $\Delta_p/k_B$ one readily
obtains the tunnelling conductance
\begin{equation}
\sigma (V, {\bf m})\equiv {dI(V, {\bf m)}\over {dV}}\propto
\sum_{\nu} | u_{\nu}({\bf m})|^2 \delta(eV-\epsilon_{\nu}),
 \end{equation}
 which is a local excitation DOS. The solution  Eq.(9) leads to a spatially modulated
 conductance,
\begin{equation}
\sigma(V, {\bf m)}= \sigma_{reg}(V) +\sigma_{mod} (V) \cos(\pi
m_x+\pi m_y).
\end{equation}
The smooth (regular) contribution is
\begin{equation}
\sigma_{reg}(V)=\sigma_0 \sum_{{\bf k},r=1,2}(u_{r{\bf k}}^2
+\tilde{u}_{r {\bf k}}^{2}) \delta(eV-\epsilon_{r{\bf k}}),
\end{equation}
and the amplitude of the modulated contribution is
\begin{equation}
\sigma_{mod}(V)= 2\sigma_0 \sum_{{\bf k},r=1,2}u_{r{\bf
k}}\tilde{u}_{r {\bf k}} \delta(eV-\epsilon_{r{\bf k}}),
\end{equation}
where $\sigma_0$ is a constant. Conductance modulations reveal a
checkerboard pattern, as the Bose condensate itself, Eq.(2),
\begin{equation}
{\sigma-\sigma_{reg}\over{\sigma_{reg}}}=A \cos(\pi m_x+\pi m_y),
\end{equation}
where
 \begin{eqnarray}
 &&A=2 \sum_{\bf k} \left[u_{1
\bf k}\tilde{u}_{1\bf k} \delta (eV-\epsilon_{1 \bf k})+ u_{2\bf
k}\tilde{u}_{2 \bf k}  \delta (eV-\epsilon_{2 \bf k})\right]/\cr
&& \sum_{\bf k} \left [(u_{1 \bf k}^2 +\tilde{u}_{1 \bf k}^2)
\delta (eV-\epsilon_{1 \bf k})+(\tilde{u}_{2 \bf k}^2+u_{2 \bf
k}^2) \delta (eV-\epsilon_{2 \bf k})\right] \nonumber
\end{eqnarray}
is the amplitude  of  modulations  depending on the voltage $V$
and temperature. An  analytical result is obtained in the
strong-coupling limit with the excitation spectrum  given by Eqs.
(16,17) for the voltage  near the threshold, $eV\approx \Delta_p$.
In this case only  states near bottoms of each excitation band,
Fig.1, contribute to the integrals in Eq.(22), so that
\begin{equation}
\tilde{u}_{1\bf k}={\xi_{\bf k}-\epsilon_{1 \bf
k}\over{\epsilon_{1 \bf k}-\xi_{\bf k-g}}}u_{1\bf k} \approx
-u_{1\bf k}{\Delta_c^2\over{ \mu w}}\ll {u}_{1\bf k},
\end{equation}
and
\begin{equation}
u_{2\bf k}={\xi_{\bf k-g}-\epsilon_{2 \bf k}\over{\epsilon_{2 \bf
k}-\xi_{\bf k}}}\tilde{u}_{2\bf k} \approx -\tilde{u}_{2\bf
k}{\Delta_c^2\over{ \mu w}}\ll \tilde{u}_{2\bf k}.
\end{equation}
Substituting  these expressions into $A$, Eq.(22), yields in the
lowest order of $\Delta_c$,
\begin{equation}
A\approx -{2\Delta_c^2\over{ \mu w}}.
\end{equation}
The result, Eq.(22) generally agrees with the STM experiments
\cite{hoff,hoff2,kapit,fu,mc}, where the spatial checkerboard
modulations of $\sigma$ were observed in a few cuprates. The
period of the modulations was found either commensurate or
non-commensurate  depending on a sample composition. In our model
the period is determined by the center-of mass wave vectors ${\bf
K}$ of the Bose-condensed preformed pairs. While the general case
has to be solved numerically \cite{alekab2}, the perturbation
result, Eq.(22) is qualitatively applied for any ${\bf K}$ at
least close to $T_{c}$, where the coherent gap is small, if one
replaces $\cos(\pi m_x+\pi m_y)$ by $\cos(K_{x} m_x+K_{y} m_y)$.
The period of DOS modulations does not depend on the voltage in
the perturbation regime, as observed \cite{ver}, but it could be
voltage dependent well below $T_c$, where higher powers of
$\Delta_c$ are important. Different from any other scenario,
proposed so far \cite{sce}, the hole density, which is about twice
of the condensate density at low temperatures, is spatially
modulated with the  period determined by the inverse wave vectors
corresponding to the center-of-mass pair band-minima. This
'kinetic' interpretation of charge modulations in cuprates,
originally proposed \cite{alesym} before STM results became
available, is consistent with the inelastic neutron scattering,
where
incommensurate inelastic peaks were observed $only$ in the $%
superconducting$ state \cite{bou}. The vanishing at $T_{c}$ of the
incommensurate  peaks is inconsistent with any other stripe
picture, where a characteristic distance needs to be observed in
the normal state as well. In our model the checkerboard charge
modulations should disappear  above $T_{c}$, where the
Bose-condensate evaporates and the coherent gap $\Delta_c(T)$
vanishes, so that $A=0$ in Eq.(22). While some  STM studies
\cite{ver} report  incommensurate DOS modulations somewhat above
$T_c$, they might be due to extrinsic inhomogeneities. In
particular, preformed pairs in the surface layer could
bose-condense at higher temperatures compared with the bulk $T_c$.
Our model is microscopically derived using the strong-coupling
(bipolaron) extension of the BCS theory \cite{ale}. If the
electron-phonon interaction is strong, such that the BCS coupling
constant $\lambda>1$, electrons form bipolarons above $T_c$, which
are Bose condensed below $T_c$. The polaron bandwidth is
exponentially reduced, which explains a low estimate of the Fermi
energy using the experimental London penetration depth in cuprates
\cite{alefermi}. Evidence for an exceptionally strong
electron-phonon interaction in high-temperature superconductors is
now overwhelming (see, for example, \cite{ZHAO,LANZ}). Yet,
generally, the model describes charge modulations  due to the Bose
condensation with non-zero center-of-mass momenta of preformed
pairs formed by any pairing interaction.

In conclusion, I solved  BdG equations with the periodic
off-diagonal potential caused by the Bose condensation  of
preformed pairs with non-zero center-of-mass momenta, and found
the checkerboard modulations of the single-particle DOS similar to
those observed in cuprates.   The main assumption
 that   the ground state of superconducting cuprates is the Bose-Einstein condensate of
 preformed pairs, is  supported by a growing number of other experiments \cite{ale}.
 The model links charge orders, pairing, and pseudo-gaps  as  manifestations of a strong attractive
 interaction in narrow bands.

The author acknowledges support of this work  by the Leverhulme
Trust (London) and by The Royal Society (UK). I would like to
thank A.F. Andreev, V.V. Kabanov, A. Lanzara, K. McElroy,   and
V.N. Zavaritsky for illuminating discussions.


\begin{thebibliography}{99}
\bibitem{sce} The  Wigner crystal of doped holes or hole
pairs is refutable because their Coulomb repulsion is relatively
weak. The relevant dimensionless parameter is $r_{s}=m^{\ast
}e^{2}/\epsilon _{0}(4\pi n/3)^{1/3}$. The Wigner crystallization
could appear at $r_{s}\gtrapprox 100$, which corresponds to an
unrealistically small atomic density of holes $n\leq 10^{-6}$ with
the static dielectric constant $\epsilon _{0}=30$ and the
effective mass $m^{\ast }=5m_{e}$.
\end{thebibliography}
\end{document}